# Child Impact Statements:
# Interdisciplinary Collaboration in Political Science and Computer Science

Leah C. Windsor[1,2]
Associate Professor of Applied Linguistics
Institute for Intelligent Systems | Department of English
The University of Memphis

Abstract

Child Impact Statements (CIS) are instrumental in helping to foreground the concerns and needs of minor community members who are too young to vote and often unable to advocate for themselves politically. While many politicians and policymakers assert they make decisions in the best interests of children, they often lack the necessary information to meaningfully accomplish this. CISs are akin to Environmental Impact Statements in that both give voice to constituents who are often under-represented in policymaking. This paper highlights an interdisciplinary collaboration between Social Science and Computer Science to create a CIS tool for policymakers and community members in Shelby County, TN. Furthermore, this type of collaboration is fruitful beyond the scope of the CIS tool. Social scientists and computer scientists can leverage their complementary skill sets in data management and data interpretation for the benefit of their communities, advance scientific knowledge, and bridge disciplinary divides within the academy.

Key terms: child impact statements; computer science; social science; collaboration; interdisciplinary

---

[1] Corresponding author: Leah.Windsor@memphis.edu

**Introduction**

In February 2008, the Shelby County Commissioners in Tennessee adopted a resolution to fund a Child Impact Statement project so that county-level decisions would be informed by the outcomes on their youngest constituents: the children . The goal of Child Impact Statements are to attach data and metrics to initiatives and proposals to demonstrate how they would benefit - and ideally not harm - children in the community. The resolution, in part, read, "A Child Impact Statement shall be submitted to the County Commission with all items presented for approval that deal with education, health, land use and any other issue affecting juvenile justice that could reasonably impact children in any substantial way." Child Impact Statements (CISs) integrate children's concerns into the policymaking process, since children cannot vote and their interests are not guaranteed to be represented at the decision-making tables.

Children are under-represented and under-counted for many reasons. Policymakers, politicians, and community members often allege they are doing things "for the children," but often provide little data or evidence to demonstrate how their efforts will benefit children. In part, this is because the data are often in disparate locations, or stored in user-unfriendly formats. In his TED talk, Hans Rosling talked about liberating data from its siloes and democratizing information so that it can be meaningfully used and interpreted (Rosling, 2006). The CIS tool for Shelby County serves as a means to aggregate and evaluate locally relevant data for children in this community in order to better inform policymaking that is responsive to their needs (Windsor, 2024). CISs are not yet widespread across the United States, but the workflow and model for initiating and enacting them is simple.

Child Impact Statements are designed to improve the quality of information available to Shelby County decision makers. Oftentimes, political decisions are influenced by more than objective information, and decisions can be influenced by implicit bias (Greenwald & Krieger, 2006), the recency effect (Henne et al., 2021), partisan preferences (Groenendyk, 2013), and historical context (Kiel, 2010, 2017). It is important, therefore, that policy-makers have a clear, data-driven picture to understand the consequences – the costs, benefits and distributional effects – of their decisions. Well-intentioned plans sometimes have unintended negative consequences, as dual-academic career couple Janukowski and Betts identified in their work on crime and housing voucher redemptions in Memphis (Rosin, 2008). Betts, a sociologist, and Janikowski, a criminologist, realized that the community networks and social capital studied by Betts were disrupted by public housing renovations, and matched nearly perfectly to the crime hot-spots identified in Janikowski's work.

In hindsight, Child Impact Statements might have helped policymakers to foresee the downstream effects of fractionalizing and dispersing communities across the county. While the goal was combating blight, the consequence was increased crime and instability for children and families. While there are national and state-level resources that could provide aggregate information for policymakers, such as the Annie E. Casey KidsCount report, and the Tennessee Commission on Children and Youth, the most useful information resides in local data and Census data. The project I supervised would marry these data streams and integrate then in a seamless and user-friendly platform to support data-informed decisions.

**Interdisciplinary collaboration**

For two years (2021-2023), I served as a mentor and PI (principal investigator) for four teams of undergraduate seniors in Computer Science in their capstone course at The University of Memphis. The goal of the capstone course for the students was to deliver a product to the mentor/PI as they would for a client in a professional setting. I presented them with an overview and motivation for the idea of Child Impact Statements, and over the course of four semesters, four different teams of capstone students helped to develop, de-bug, and deploy an interactive website that contextualizes children's well-being in Shelby County.

We met weekly for an hour via Zoom each semester. The semester was divided into four Sprints, a term used to delineate a time frame with specific goals and deliverables for the students to meet. I submitted a Qualtrics form evaluation of the students' performance for each Sprint to provide the instructor of record with qualitative information about their work. The workflow and code for the CIS tool is stored in a Github repository.

Sprint 1 began with a listening session where I described my vision for the project. During this Sprint I also collected the URLs and APIs for the students to integrate into the CIS tool. We decided it would be a browser-based tool that pulled updated information from publicly available sources, as well as some static data from the Shelby County Public Health Department on COVID-19 vaccinations and elevated blood lead levels that could be manually updated as needed. The students provided a "wire frames" overview of the tool, the equivalent of an architect's visual rendering of blueprints for a building project.

In Sprint 2, the students started connecting the data pipelines and integrating them into a geospatial map-based format. In Sprint 3, the students solicited feedback from me about the usability of the website, including font, colors, icons, and display. In Sprint 4, the students did a "think aloud" process with confederates who completed a series of instructions and provided feedback in real time. The CIS tool was ready to deploy at the end of the first semester. Students in subsequent semesters improved upon various features of the tool, including its functionality and visual and aesthetic appearance.

Users can choose between retrieving data at the zip code, Census tract, or commissioner district level (Figure 1). Data is presented in a sidebar when users click on the geographic quantity of interest, and may be downloaded as a .pdf or in .csv formats. Graphs for some of the data are available for download as well (Figure 2). Figure 3 shows elevated blood lead levels in children across Shelby County, disaggregated by zip code. Lead exposure can come from lead pipes in older building structures, lead paint, and soil contamination. There is no acceptable or tolerated level of lead exposure in children, and elevated blood lead levels are associated with cognitive and learning problems in children (Edwards, 2013; O'Connor et al., 2020; Wright et al., 2021). Using the available data found in the CIS tool, policymakers can propose lead remediation and interventions in places where blood lead levels are highest. In effect, the CIS tool is a precision instrument that can show policymakers and the community where the greatest return on their investments will be felt.

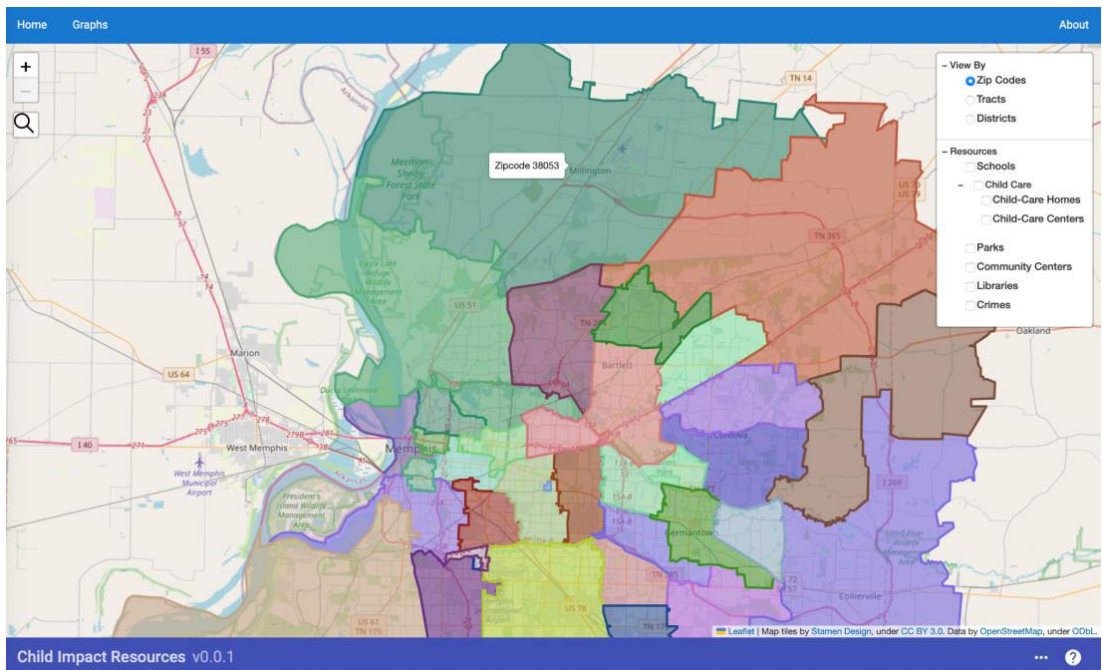

**Figure 1. Child Impact Statement Tool, http://ShelbyCountyKidData.com**

**Figure 2. Child Impact Statement Tool pre-formatted graphs and maps for download**

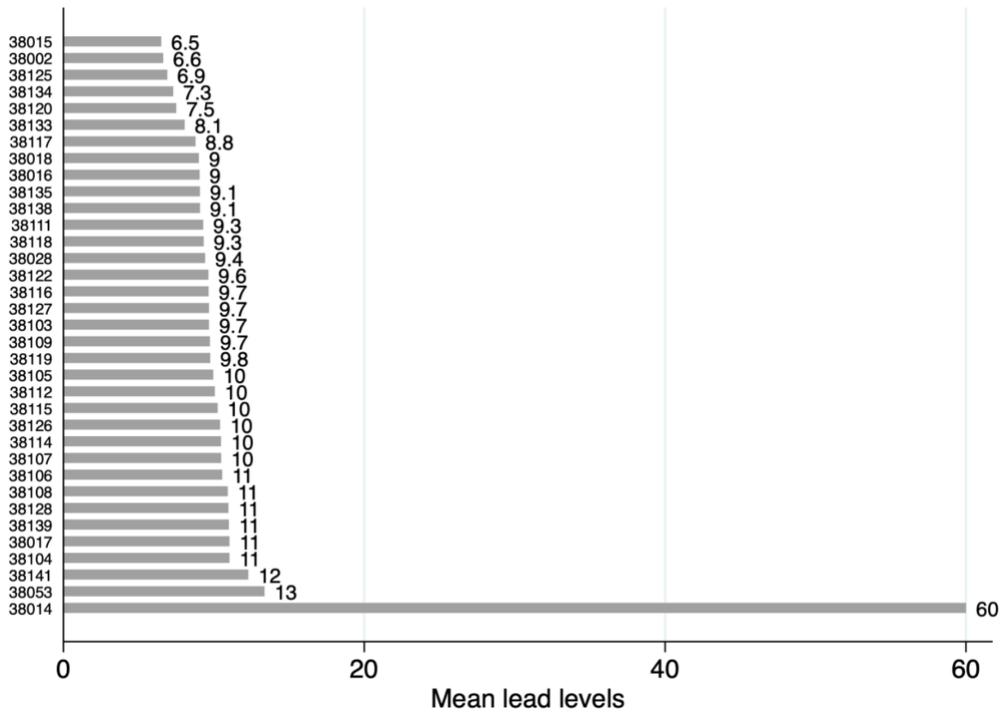

**Figure 3. Elevated blood lead levels by zip code**

**Lessons learned**

As a computational social scientist, I have the opportunity to work across many fields and with scholars from many disciplines. The challenges of interdisciplinarity - including learning new vernacular, methodologies, and literatures - are far outweighed by the benefits - in this case leveraging the strengths of computer scientists in coding and programming to develop a social-science informed tool for improving public policymaking. The following are recommendations for interdisciplinary collaborations:

1. Make your expectations clear and use jargon-free language to communicate your ideas.
2. Have three "streams" of code: programming code, "commented" code, and plain language translation of the programming code. The computer scientists who write the programming code should also "comment" (delineated by // in the data files) on what each segment of code does. The social scientist can work with the computer scientists to write the plain language description of the code and comments.
3. For the mentor/PI, include students in the policymaking process with local elected officials and community members. Service learning is increasingly important in connecting students with future employment opportunities, and demonstrating the value of higher education degrees.

The CIS tool itself, and the interdisciplinary, collaborative development process that begat it, can be replicated in other universities and communities. The tool should be useful for many stakeholders, and provides evidence for use-inspired learning and service-oriented education.

**Works Cited**

Edwards, M. (2013). Fetal death and reduced birth rates associated with exposure to lead-contaminated drinking water. *Environmental Science & Technology*, *48*(1), 739–746.

Greenwald, A. G., & Krieger, L. H. (2006). Implicit Bias: Scientific Foundations. *California Law Review*, *94*(4), 945–967. https://doi.org/10.2307/20439056

Groenendyk, E. (2013). *Competing Motives in the Partisan Mind: How Loyalty and Responsiveness Shape Party Identification and Democracy*. OUP USA.

Henne, P., Kulesza, A., Perez, K., & Houcek, A. (2021). Counterfactual thinking and recency effects in causal judgment. *Cognition*, *212*, 104708. https://doi.org/10.1016/j.cognition.2021.104708

Kiel, D. (2010). A Memphis Dilemma: A Half-Century of Public Education Reform in Memphis and Shelby County from Desegregation to Consolidation. *U. Mem. L. Rev.*, *41*, 787.

Kiel, D. (2017). Exploded Dream: Desegregation in the Memphis City Schools. *Law & Inequality: A Journal of Theory and Practice*, *26*(2), 261.

O'Connor, D., Hou, D., Ok, Y. S., & Lanphear, B. P. (2020). The effects of iniquitous lead exposure on health. *Nature Sustainability*, *3*(2), 77–79. https://doi.org/10.1038/s41893-020-0475-z

Rosin, H. (2008, August). American Murder Mystery. *The Atlantic*. https://www.theatlantic.com/magazine/archive/2008/07/american-murder-mystery/306872/

Rosling, H. (Director). (2006). *The best stats you've ever seen*. https://www.ted.com/talks/hans_rosling_shows_the_best_stats_you_ve_ever_seen

Windsor, L. (2024). *Child Impact Statements*. Child Impact Statements. https://shelbycountykiddata.com/

Wright, J. P., Lanphear, B. P., Dietrich, K. N., Bolger, M., Tully, L., Cecil, K. M., & Sacarellos, C. (2021). Developmental lead exposure and adult criminal behavior: A 30-year prospective birth cohort study. *Neurotoxicology and Teratology*, 106960. https://doi.org/10.1016/j.ntt.2021.106960